\documentclass[11pt]{article}
\pdfoutput=1
\usepackage{jcappub} 
\usepackage{cancel}
\usepackage[T1]{fontenc} 
\usepackage{tikz-feynman}
\usepackage{amsmath}
\usepackage{graphicx}

\newcommand{\beq}{\begin{equation}\begin{aligned}{}}
\newcommand{\eeq}{\end{aligned}\end{equation}}
\newcommand{\beqa}[1]{\begin{equation}\begin{aligned}{#1}}
\newcommand{\eeqa}{\end{aligned}\end{equation}}

\newcommand{\bea}{\begin{eqnarray}{}}
\newcommand{\eea}{\end{eqnarray}}

\newcommand{\nn}{\nonumber}

\newcommand{\del}{\partial}

\usepackage{slashed}
\usepackage{algorithm}

\title{\boldmath Anomaly induced cooling of Neutron Stars: A Standard Model contribution}
\author[a]{Sabyasachi Chakraborty,}
\author[b]{Aritra Gupta,}
\author[c]{Miguel Vanvlasselaer}

\affiliation[a]{Department of Physics, Indian Institute of Technology, Kanpur-208016, India}
\affiliation[b]{Instituto de F\'isica Corpuscular (IFIC), CSIC,
Parc Cient\'ific, C/Catedr\'atico Jos\'e Beltr\'an, 2, E-46980 Paterna, Spain}
\affiliation[c]{Theoretische Natuurkunde and IIHE/ELEM, Vrije Universiteit Brussel,
\& The International Solvay Institutes, Pleinlaan 2, B-1050 Brussels, Belgium}

\emailAdd{sabyac@iitk.ac.in}
\emailAdd{aritra.gupta@ific.uv.es}
\emailAdd{miguel.vanvlasselaer@vub.be}

\abstract{
Young neutron stars cool via the emission of neutrinos from their core. A precise understanding of all the different processes producing neutrinos in the hot and degenerate matter is essential for assessing the cooling rate of such stars. The main Standard Model processes contributing to this effect are $\nu$ bremsstrahlung, mURCA among others.  In this paper, we investigate another Standard Model process initiated by the Wess-Zumino-Witten term, leading to the emission of neutrino pairs via $N\gamma\to N\nu\bar\nu$. We find that for proto-neutron stars, such processes with degenerate neutrons can be comparable over the typical and well-known cooling mechanisms only if the coupling $g_\omega \gtrsim 20$.}

\subheader{IFIC/23-23}

\begin{document}
\maketitle
\flushbottom

\section{Introduction}
Neutron stars (NS) are one of the most enigmatic celestial objects present in our Universe. Due to its very high density and large magnetic field neutron stars provide extreme laboratory conditions to test the existing laws of physics and hunt for new ones which would otherwise be impossible to study in terrestrial experiments. Apart from being a testing field for new physics, NS are interesting objects of study in their own accord. Because of extremely dense cores, various states of compressed nuclear matter can co-exist most probably in a superfluid neutron degenerate state~\cite{Page:2013hxa}. The core may also contain exotic states of matter like strange matter and quark-gluon plasma. However, the exact composition of the core is not yet known~\cite{Sedrakian:2006mq}\footnote{For reviews on the composition \cite{1983bhwd.book.....S, Glendenning:1997wn, Blaschke_Glendenning_Sedrakian_2001, Haensel:2007yy} and cooling of NS see~\cite{Raffelt:1996wa,Yakovlev:2000jp}.}. Theoretical understanding of the core is difficult due to our dearth of knowledge regarding the theory of strong nuclear interactions and the exact many-body theory. The only insight into the inner structure of the NS is via experimental signatures. One can try to constrain theoretical predictions of different equations of state by measuring the stellar mass and/or the radius. Another way to investigate the inner composition of a NS is by studying its cooling mechanism and comparing theoretical predictions with the observed luminosity. A NS first cools itself mostly by emitting neutrinos and in the later stage, when $T \lesssim 10^8$ K ($0.01$ MeV), by emitting photons mostly from its surface. Since the neutrinos interact extremely weakly, they are able to stream freely even out of such a dense environment unless the neutron stars are extremely hot and are in their proto-neutron phase. Neutron stars are transparent to neutrinos as long as their temperature is $\lesssim \mathcal{O}\,(10)$ MeV~\cite{2306.13591, PhysRevD.67.123002, Camelio:2017hrs}. Different cooling mechanisms dominate depending on whether the process is taking place in the crust or near the core~\cite{Yakovlev:2000jp}. It also depends on the temperature and the age of the NS. During the first 20 s of the NS, cooling occurs predominantly by pair production of neutrinos through Bremsstrahlung \cite{fischer2012neutrino}, URCA and mURCA among other processes~\cite{PhysRevLett.66.2701}.

In this paper, we investigate in detail a new way of producing $\nu\, \Bar{\nu}$ inside NS without introducing any new beyond Standard Model (SM) physics. This new production channel can thus in turn act as an additional mode of cooling of the NS. This channel originates from the Wess-Zumino-Witten (WZW) term which was originally introduced to account for processes allowed by QCD but apparently forbidden by the spurious parity and charge conjugation symmetries of the effective Chiral Lagrangian \cite{Wess:1971yu, Witten:1983tw, Kaymakcalan:1983qq, Chou:1983qy, Kawai:1984mx, Pak:1984bn, Harvey:2007ca}. As an example, the WZW term helps us to explain observed processes like $\pi^0 \to \gamma \gamma$ and  $K^+\, K^- \to \pi^+\,\pi^-\pi^0$, which are otherwise forbidden by the symmetries of the Chiral Lagrangian. Gauging this WZW term leads to effective vertices connecting the $U(1)_{\rm em}$ photon, $SU(2)$ gauge boson, and the vector mesons. For example, this term contains an interaction of the form $ \epsilon^{\mu \nu \alpha \beta} \,F_{\mu \nu} \,\omega_\alpha \,Z_\beta$ where $Z$ is the Standard Model $Z$ boson, $\omega$ is the omega meson and $F_{\mu \nu}$ is the usual SM $U(1)_{\rm em}$ field strength. This term is thus an unavoidable consequence of the SM anomaly~\cite{Harvey:2007ca}.

The importance of such a term in the cooling mechanism of a NS was first studied in \cite{Harvey:2007rd} where the authors showed that a $U(1)_{\rm em}$ photon, after getting a mass due to the medium effects inside the superconducting core of a NS, can decay into two neutrinos, thereby contributing to the cooling process of the star. However, the analysis lacked a proper treatment of the degeneracy of the strongly packed neutrons. In the core of the NS, neutrons are in a degenerate phase and this can severely constrain the allowed phase space for any cooling mechanism involving by-standing neutrons. In this work, we take into account the neutron degeneracy in a formal way and calculate the emissivity of the process $N\,\gamma \rightarrow N\, \nu\,\Bar{\nu}$ where the $N$ in the initial and the final state represent by-standing neutrons (Fig\, \ref{fig:Cooling_diag}). This is analogous to the mURCA process where by-standing neutrons are introduced in the initial and the final state to conserve energy and momentum. The mURCA is the dominant cooling channel when the usual URCA process becomes inactive below a critical density of 3\,$\rho_0$ where $\rho_0 \sim 3\times 10^{14}$ g cm$^{-3}$. We derive analytical results for the emissivity of neutrinos via the process shown in Fig\, \ref{fig:Cooling_diag}. At temperatures lower than $m_\omega$ and $M_Z$, we integrate out the heavy $Z$ and $\omega$, leading to an effective interaction involving only neutrinos, photons, and neutrons. We find that the parametric dependence of the emissivity on temperature and mass of the photon is different from what the authors of \cite{Harvey:2007rd} obtained in their earlier work. Furthermore, we also note that a simplistic way to incorporate the degeneracy as an overall multiplicative factor of $F_{\rm Deg} \sim T/M_N$~\cite{Raffelt:1996wa}, deviates from the actual calculation by several orders of magnitude depending on the temperature of the NS ($T$).

The rest of the paper is organised as follows:
In section \ref{section:SMproc} we review different cooling channels of NS from the Standard Model dominant processes, in section \ref{sec:lagrangian}, we introduce the Lagrangian of the WZW term, in section \ref{sec:neutrino_HHH} we present the computation of the cooling rate due to this process including carefully the effect of Pauli blocking from neutron degeneracy, and then compare our results with those previously found in the literature. Finally in Section \ref{sec:conclusion}, we conclude and discuss some future possibilities.

\section{Usual Standard Model processes contributing to the cooling of a NS}
\label{section:SMproc}
In this section, we briefly discuss some usual SM processes that contribute to the cooling of a hot NS. In the crust of a NS, characterised by a density typically of the order of $\sim 0.5 \,\rho_0$, one of the most important cooling channels is the plasmon decay. The plasmons can be thought to be photons that have gained mass due to in-medium effects via electron loops~\cite{PhysRev.129.1383,PhysRevD.40.3679,PhysRevD.67.123002}. The ultra-relativistic and degenerate electrons inside the NS constantly scatter with the photons resulting in a deformation of the dispersion relation for those photons propagating inside the NS. The process is most dominant as long as the plasma frequency $\omega_{\rm pe} \lesssim T$ and scales as $\rho^2$~\cite{Yakovlev:2000jp} whereas for lower temperatures it is suppressed exponentially. Computation of electron loops in a degenerate medium results in $\omega_{\rm pe} \sim m_{\gamma}\sim \sqrt{4\pi e^2\,n_e/\mu_e}$, where $\mu_e$ ($n_e$) is the chemical potential (density) of the electrons in the NS~\cite{Raffelt:1996wa}. For a hot NS with temperatures around a MeV, $m_{\gamma}\sim 1-10$ MeV depending on the equation of state of the NS~\cite{Raffelt:1996wa,Yakovlev:2000jp,Voskresensky_1998}. Other competing processes at the crust are electron-nucleus Bremmstrahlung~\cite{1989ApJ...339..354I} ~($e(A,Z) \rightarrow e(A,Z) \,\nu \Bar{\nu}$), electron synchrotron radiation~\cite{1997A&A...328..409B} ($e\rightarrow e \nu \Bar{\nu}$) and electron positron annihilation\,\cite{1967ApJ...150..979B,1989ApJ...339..354I,PhysRevD.6.941} ($e^+ e^-\to \nu\bar\nu$).

From the crust, as we proceed towards the centre of the NS, a plethora of different cooling mechanisms open up. In the outer core characterised by a density of $0.5\,\rho_0 \lesssim \rho \lesssim 3\, \rho_0$, the most important cooling mechanism is the modified URCA process~\cite{1995A&A...297..717Y,1979ApJ...232..541F}, symbolically denoted by,
\bea
n \, n \longrightarrow n \, p \, e\, \Bar{\nu_e}\;,\;\;\;\;
n \, p\,e \longrightarrow n \, n \, \nu_e\;.
\eea
This can also be followed by an analogous proton branch. The emissivity for the said reaction is given by~\cite{Yakovlev:2000jp}
\bea
Q^{\rm mURCA} \simeq 10^{26-29}\,\left(\dfrac{T}{1\,\rm MeV}\right)^8\,\rm erg \, s^{-1} \,cm^{-3}\;.
\label{eq:mURCA}
\eea
The uncertainty in the order of magnitude mainly results from in-medium effects. In principle, the value of the weak-interaction couplings that enter the calculation should be renormalised due to the medium effects. Other sources of uncertainties may lie in the discrepancy of the pion coupling and other nuclear physics factors. In the outer core, another process that also assists in cooling is via the bremsstrahlung of neutrinos and anti-neutrinos in a baryon-baryon collision. These are neutral current interactions. In ordinary nuclear matter, the energy-loss rate for this process is a little slower than the mURCA. The emissivity is given by~\cite{Yakovlev:2000jp,1979ApJ...232..541F,1995A&A...297..717Y}
\bea
Q^{\rm \,\nu-Brem} \simeq 10^{24-28}\,\left(\dfrac{T}{1\,\rm MeV}\right)^8\,\rm erg \, s^{-1} \,cm^{-3}\;.
\label{eq:nubrem}
\eea
Inside extremely hot ($\sim 10$ MeV) NS at around a density of $\rho_0$, cooling may also take place via photoneutrino process, i.e., $e^\pm \gamma \rightarrow e^\pm \nu \Bar{\nu}$~\cite{PhysRevD.69.023005} and via decay of a plasmon mode~\cite{PhysRevD.67.123002}. Similar plasma excitation of a $pe\mu$ three fluid system at similar densities have been studied in \cite{Baldo:2009we}.

In addition to the normal $npe\mu$ nuclear matter, the outer core may also consist of baryons in a superfluid state where the baryons form Cooper pairs provided $T < T_c$ with $T_c$ being the critical temperature of the system. Such cooper pairs can spontaneously break $U(1)_{\rm em}$ giving rise to a massive photon. Within the superfluid, the mass of the photon is given by $m_\gamma \sim \sqrt{8\,\pi\,e^2\,\phi_c^2}$ where $\phi_c$ is the order parameter of the system. For proton cooper pairs, it is shown in \cite{Voskresensky_1998} that $m_\gamma \sim 1.6\,\sqrt{(T_c-T)/T_c}$ MeV around a density of $\rho_0$. For such cooper pairs in a typical NS $T_c \sim \mathcal{O}(\rm MeV)$. This implies that typical masses of such photons will be $\lesssim$ MeV. However, for Kaon or charged pion condensates or for di-quark condensates $T_c$ can be as large as 50 MeV~\cite{PhysRevLett.81.53,Voskresensky_1998}. These photons can decay to neutrinos and help in the cooling of a hot NS via the Pair Breaking and Formation (PBF) mechanism \cite{Flowers:1976ux,Voskresensky_1998}. The emissivity for such a process is given by~\cite{Voskresensky_1998}
\bea 
Q^{\rm \,PBF} \simeq  10^{27} \bigg(\frac{T}{\text{MeV}}\bigg)^{3/2} e^{-m_\gamma/T}  \bigg(\frac{m_\gamma}{\text{MeV}} \bigg)^{7/2}  \bigg(\frac{\rho}{\rho_0}\bigg)^{8/3}   \bigg(1+ \frac{3}{2}\frac{T}{m_\gamma} \bigg) (1+ \eta) \,\rm erg \, s^{-1} \,cm^{-3}
\eea 
where $\eta$ is a small parameter related to the pairs of proton holes. Cooling of NS by a decaying massive photon can also occur via the WZW term, first studied briefly in~\cite{Harvey:2007rd}. Such cooling processes will be the main focus of study in the rest of this paper.

If we continue moving inwards to regions of density $\sim 4\, \rho_0$, another important cooling mechanism called Direct URCA becomes dominant over the mURCA\cite{PhysRevLett.66.2701, 1992ApJ...390L..77P}. The process is similar to the mURCA but without the spectator neutrons in the initial and final state,
\bea
n \longrightarrow  p \, e\, \Bar{\nu_e}\;,\;\;\;
p\,e \longrightarrow n \, \nu_e\;.
\eea
The by-standing neutrons in the mURCA help to conserve the four-momentum and allow the process at lower densities as compared to the direct URCA. Consequently, the emissivity in the case of direct URCA scales as $T^6$ with respect to the NS temperature. 

In a light NS, i.e., when $M_{\rm NS} \lesssim M_{\rm sun}$, the density threshold for the activation of the fast URCA processes is never reached~\cite{Blaschke:2011gc, Hess:2011qw} and cooling is significantly slowed down. This is the so-called \emph{minimal cooling} paradigm~ \cite{2004ApJS..155..623P,2009ApJ...707.1131P}. Neutron stars where the direct URCA process is dominant are said to experience \emph{enhanced} cooling in the core. The latter effect is propagated throughout the volume of the star and as a consequence, the surface temperature also falls off quite rapidly in such cases. Thus, processes which are slower than the URCA like the mURCA, will be less effective in cooling the star. However, even at high densities, the activation of direct URCA is not guaranteed and may also depend on the nuclear physics models at hand~\cite{2306.13591}. In this paper, we assume a minimal cooling regime.

On top of the SM cooling processes, many beyond SM processes can also lead to an enhanced cooling of the NS. New light particles with sizeable coupling to the SM would lead to new channels of cooling in the NS which could lead to discrepancies between the observation of the cooling curve and SM based simulations. The existence of an axion induces almost unavoidably a coupling to the nucleons (see \cite{DiLuzio:2017ogq, Badziak:2023fsc, Takahashi:2023vhv} for an exception)  and an enhanced cooling of NS that permits putting bounds on the axions parameters \cite{PhysRevLett.53.1198,Raffelt:1996wa,Umeda:1997da, Weber:2006ep, Leinson:2014ioa, Sedrakian:2015krq, Buschmann:2021juv} and for proto-NS in \cite{Leinson:2021ety,Fischer:2021jfm}. Dark gauge bosons \cite{Hong:2020bxo, Shin:2021bvz} couplings can also be constrained on the basis of similar arguments.

\section{Road to the Lagrangian} 
\label{sec:lagrangian}
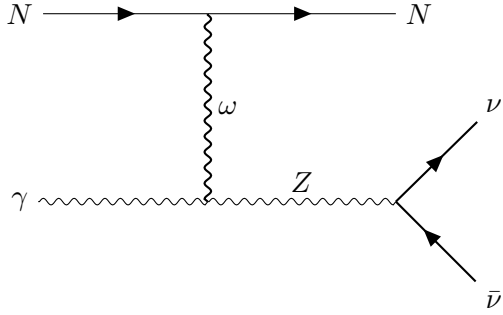
\begin{figure}[h!]
\center
\begin{tikzpicture} 
\begin{feynman}
\vertex (a1) {\(N\)}; 
\vertex[right=2.5cm of a1] (a2); 
\vertex[right=2.5cm of a2] (a3) {\(N\)}; 
\vertex[below=2.5cm of a1] (b1) {\(\gamma\)}; 
\vertex[right=2.5cm of b1] (b2) ;
\vertex[right=2.5cm of b2] (b3); 
\vertex[above right=1.5cm of b3] (g){\(\nu\)}; 
\vertex[below right=1.5cm of b3] (h){\(\bar\nu\)}; 
\diagram*{
(a1) -- [fermion] (a2),
(a2) --[fermion] (a3),
(b1)--[boson] (b2),
(b2)--[boson, edge label= {\(Z\)}] (b3),
(a2)--[boson, thick, edge label=\(\omega\)](b2),
(b3)--[fermion, thick](g),
(b3)--[anti fermion, thick](h),
};
\end{feynman} 
\end{tikzpicture}
 \caption{Neutron star cooling due to WZW interactions. Feynman diagram of the process $N \gamma \to N \nu \nu$ inducing the new channel for the neutron star cooling.}
\label{fig:Cooling_diag}
\end{figure}

As mentioned in the introduction, we are interested in the process $N\gamma\to N\nu\bar\nu$ shown in Fig.~\ref{fig:Cooling_diag}. Firstly, Dirac spinors or in our case, nucleons interact with the meson fields such as $\omega$ via
\begin{equation}
\mathcal L_0 = \bar{N}\left(i\cancel{\partial}-g_\omega \cancel{\omega}-M_N\right) N\;.
\label{eq:1st_Lag}
\end{equation}
$g_\omega$ is the coupling constant or form factor. 
The numerical value depends on the model\footnote{We thank Andrea Caputo for useful discussions about the experimental determination of the value of $g_\omega$. }. Authors in~\cite{Downum:2006re} suggest that $g_\omega$ is related to the pion-nucleon coupling via $
    g_\omega = 9/5\left(m_\omega/m_\pi\right)^{1/2} g_\pi, 
    $, which suggests values between $30-60$. Phase shift data and scattering data for energy range $E \in [25-500]$ MeV however suggest a value $g_\omega \sim 10$ (see the PIONS@MAX-lab Collaboration~\cite{Strandberg:2018djk} results and\cite{Briscoe:2020qat,A2:2019yud,Drechsel:2007if}.), in this range of energies. We emphasize that the energy scale of the photo-production studied in this paper is however lower $E \lesssim 10$ MeV. We will thus report the cooling for different values of $g_\omega$.

However, to generate the process shown in Fig.~\ref{fig:Cooling_diag}, we would also require a coupling between $\omega-\gamma-Z$ boson~\cite{Harvey:2007ca,Harvey:2007rd}. Such interactions are generated while gauging the WZW term. WZW terms are important as the Chiral Lagrangian fails to capture certain important physics pertaining to the interaction of mesons. The terms in the Chiral Lagrangian are invariant under the quark flavor symmetries $SU(3)_L\times SU(3)_R$ and manifest more symmetries, such as the spurious parity symmetry, which is absent in the UV theory, i.e., QCD. We note in passing that it is often the case for many other effective theories such as HQET, SCET, etc. WZW term lifts this symmetry and therefore helps to mediate processes such as $K\bar K\to 3\pi$ etc. However, gauging the WZW term should be done carefully. One can gauge the anomaly-free subgroup i.e., $U(1)_{\text{em}}$ of the flavor symmetries, leading to the correct description for the $\pi^0\to\gamma\gamma$ process. On the other hand, gauging an arbitrary subgroup of the chiral symmetry group, for example, $SU(2)_L\times U(1)_Y$ which resides in the non-diagonal subgroup of the former, is rather subtle. The reason is twofold: Firstly, one has to make sure that the anomalies between the Chiral Lagrangian and lepton sectors cancel to obtain an anomaly-free theory. Secondly, introducing mesons, such as $\omega$ in the form of a background gauge field would introduce mixed anomalies, requiring new counterterms. This leads to  the following interactions~\cite{Harvey:2007ca} (see Appendix \ref{app:WZWterm} for more details):
\begin{equation}
    \mathcal L_{\text{WZW}} \supset \frac{N_C}{48\pi^2} g_2^2 g_\omega\;\tan\theta_W\;\epsilon^{\mu\nu\rho\sigma}\; F_{\mu\nu}\omega_\rho Z_\sigma\; + ...
    \label{eq:WZW_Lag}
\end{equation}
A Combination of both Eq.~(\ref{eq:1st_Lag}) and (\ref{eq:WZW_Lag}) generate our desired process $N\gamma\to N\nu\bar \nu$, as explained in the next section. 

\section{Cooling of NS via the anomaly mediated process}
\label{sec:neutrino_HHH}

In this section we consider cooling of a hot NS via the process shown in Fig.\ref{fig:Cooling_diag} after integrating out the heavy $\omega$ and $Z$. The two neutrinos escape from the NS carrying away a part of its internal energy and thereby contributing to the cooling. A similar process involving a pion as the mediator can in principle also contribute to the cooling process. However as mentioned in~\cite{Harvey:2007rd}, terms $\propto \epsilon_{\mu\nu\rho\sigma}\,\del_{\mu}\pi^0\,Z^\nu\,F^{\rho\sigma}$ would be suppressed by $f_\pi^4$, and $g_\pi/f_\pi < g_\omega/m_\omega$. Therefore, in the rest of the paper, we will not discuss the pion or other vector meson-mediated processes. The calculation of the emissivity in such a case will however be more or less analogous to what is presented here.

\subsection{The scattering matrix of $\gamma\,N \rightarrow \gamma\,N \nu \Bar{\nu}$ with $\omega$ exchange}
\label{subsec:M}

The scattering matrix of the reaction $N(p_{N_1})+ \gamma(p_\gamma) \to N(p_{N_2}) + \nu(p_1)+ \bar\nu(p_2) $ is given by
\bea 
\mathcal{M} = \kappa\;\epsilon_{\mu \alpha \rho \sigma}\; \left[ \bar \nu \gamma^\alpha (1 -\gamma_5)\nu  \; \bar N \gamma^\mu N \right]\;p_\gamma^{\rho} \epsilon^\sigma\;, \; \; \; \text{where}\; \; \; \kappa = \frac{N_c}{12\pi^2}\frac{g_\omega^2}{ m_\omega^2}\frac{e\,G_F}{\sqrt{2}}\;.
\eea
$\epsilon^\sigma$ is the polarisation vector of the photon. All the other symbols have their usual meaning. 
Next, we compute the square of the above amplitude. As we will see soon, due to the high degeneracy the momentum flowing in the $\omega$ propagator is controlled by $T \ll m_\omega$ and consequently we can safely integrate out the heavy $\omega$. In the same manner, we also integrate out the $Z$ boson. Further, we also average over the spin and polarization of the incoming particles and sum over the spins of the outgoing ones. The squared average amplitude is thus given by
\begin{eqnarray} 
\langle |\mathcal{M}|^2\rangle = \frac{1}{2 \times 3}\sum_{s, s', \lambda} \mathcal{M}\mathcal{M}^\dagger 
    =- \frac{\kappa^2}{6}\text{Tr}\left[(\slashed{p_{N_1}} + M_N)\gamma^\alpha (\slashed{p_{N_2}} + M_N)\gamma^\beta \right] \times \nonumber \\
\text{Tr}\left[\gamma^\sigma (1-\gamma_5) \slashed{p}_2\gamma^\rho (1-\gamma_5)\slashed{p}_1\right] \epsilon_{\alpha \sigma \nu \eta} \,\epsilon_{\beta \rho \mu \eta} \,p_{\gamma }^\nu \, p_{\gamma }^\mu\;.
\end{eqnarray}
In Sec.~\ref{section:SMproc} we discussed that medium effects can give rise to photons with mass in the range of 1-10 MeV. Also, neutron stars are transparent to neutrinos as long as $T \lesssim \mathcal{O}\,(10)$ MeV.
Therefore, since these photons are non-relativistic and approximately at rest with respect to the neutrons, the outgoing neutrinos are emitted almost back to back. Retaining only the dominant contributions in the limit when $M_N \gg |\vec{p}_{N_1,N_2}|$ and $m_\gamma \gtrsim |\vec{p}_{\gamma}| $, we obtain
\bea 
\langle |\mathcal{M}|^2\rangle  \approx \dfrac{64}{3} \kappa^2 M_N^2 \,E_1 \,E_2
\, p_\gamma^2 \,\bigg(1+ c_1^2+ \underbrace{\frac{1}{4}\frac{p_\gamma^2}{m_\gamma^2}}_{\text{sub-leading}}\bigg)\;.
\label{eq:mod_M_s}
\eea 
where $c_1$ is the cosine of the angle between the first neutrino and the photon. Note, $0 \leq c_1^2 \leq 1$. Conservatively we thus have,
\bea 
\langle |\mathcal{M}|^2\rangle  \approx \dfrac{64}{3} \kappa^2 M_N^2 \,E_1 \,E_2  \, p_\gamma^2\;.
\label{eq:Msqr}
\eea

\subsection{Computation of the cooling process with Pauli blocking effect}
The rate of energy released per unit volume of the neutron star material is measured by its emissivity which is defined as
\begin{eqnarray}
Q^{2\to 3} &\equiv& n_F\,\int \frac{d^3 p_\gamma}{(2\pi)^3 2E_\gamma} {g_\gamma}f_\gamma (p_\gamma) \int \frac{ {g_{N_1}}d^3 p_{N_1}d^3 p_{N_2}}{(2\pi)^3 2E_{N_1}(2\pi)^3 2E_{N_2}} \int \frac{d^3 p_{1}d^3 p_{2}}{(2\pi)^3 2E_{1}(2\pi)^3 2E_{2}} \underbrace{(E_1+ E_2)}_{\text{escaping energy}} \nonumber \\
&\times& \langle |\mathcal{M}|^2\, \rangle (2\pi)^4 \, f_N(E_{N_1})\,(1-f_N(E_{N_2}))\nonumber \\
&\times& \,\delta(E_{N1}+Q_0-E_{N_2})\delta^3(\vec p_{N_1}-\vec p_{N_2}+\vec q)\,\Theta(E_{N_1}-M_N) \,\Theta(E_{N_2}-M_N)\;,
\label{eq:Def:cooling_rate}
\end{eqnarray}
where $n_F$ is the number of families of neutrinos. Also, $p_\gamma,\,p_{N_1},\,p_{N_2},\,p_1,\,p_2$ represents the momentum of the photon, the neutron in the initial and final state, and the two emitted neutrinos respectively. Similarly, $E_{N_1}$, $E_{N_2}$, $E_1$ and $E_2$ are the energies of the two by-standing neutrons in the initial and final state and the two outgoing neutrinos respectively. We also defined $q^\mu \equiv (Q_0, \vec {q}) = p^\mu_\gamma-p^\mu_{1}- p^\mu_{2} $. 

The expression for emissivity can be written more transparently in terms of neutron response functions. This way of decomposition helps us to separate the effect of by-standing fields of highly degenerate neutrons from that of the incoming and outgoing particles. Upon using the expression for the matrix element squared from Eq.\eqref{eq:Msqr}, the emissivity in terms of the response function, $S(q)$, takes the form
\bea
Q^{2\to 3} =\dfrac{64\,n_F}{4}\frac{g_\gamma}{3} \kappa^2 \int \dfrac{d^3p_\gamma}{(2\pi)^3}
\dfrac{ f_\gamma}{2E_\gamma}|\vec{p}_\gamma |^2
\int \dfrac{d^3p_{1}\,d^3p_{2}}{(2\pi)^6 2E_{1}\,2E_{2}}  E_{1}\,E_{2} \, (E_{1}+E_{2}) \, S(q^\mu)\;,
\label{eq:mainQdef}
\eea
where the nuclear response function $S(q^\mu)$ is given by,
\begin{align}
S(q^\mu) \equiv & g_{N_1}  \int \dfrac{d^3p_{N_1}d^3p_{N_2} M_N^2}{(2\pi)^6\, E_{N_1}\,E_{N_2}} f_N(E_{N_1})(1-f_N(E_{N_2})) \times (2\pi)^4 \times \nonumber \\
&\delta(E_{N1}+Q_0-E_{N_2})\delta^3(\vec p_{N_1}-\vec p_{N_2}+\vec q)\,\Theta(E_{N_1}-M_N) \,\Theta(E_{N_2}-M_N)\;.
\label{eq:SDEF2}
\end{align}
Such response functions were previously calculated for $2 \leftrightarrow 2$ processes~\cite{Reddy:1997yr,Bertoni:2013bsa,Bell:2021fye,Garani:2020wge,Garani:2018kkd}. We outline the full calculation for our case in Appendix~\ref{app:computation}. The final form of the response function simplifies to:
\bea
S(Q_0, q)=\dfrac{M_N^2\,T}{\pi q}\dfrac{z}{1-e^{-z}}\Theta(\mu - E_-)\;.
\label{eq:mainS}
\eea
where $z \equiv  Q_0/T$ and $q \equiv |\vec{q}|$ and we have used $g_{N_1}=2$. 
The response function helps to manifestly highlight the effect of degeneracy suppression of the reaction rates in highly Fermi degenerate matter via the $\Theta$ function. In highly dense and degenerate systems, the fermions tend to completely occupy all the lower energy states and therefore it is very difficult to change the total number of particles from this sea of fermions. In other words, Fermi-degenerate matter like neutron stars are characterised by large chemical potentials. All of this is encapsulated in the $\Theta$ function which forces the minimum kinetic energy of the neutron in the initial state, i.e., $E_-$, to be smaller than the chemical potential of the NS system. If not, the reaction rates will be exponentially suppressed.

\begin{figure}[!h]
 \centering
 \includegraphics[scale=0.7]{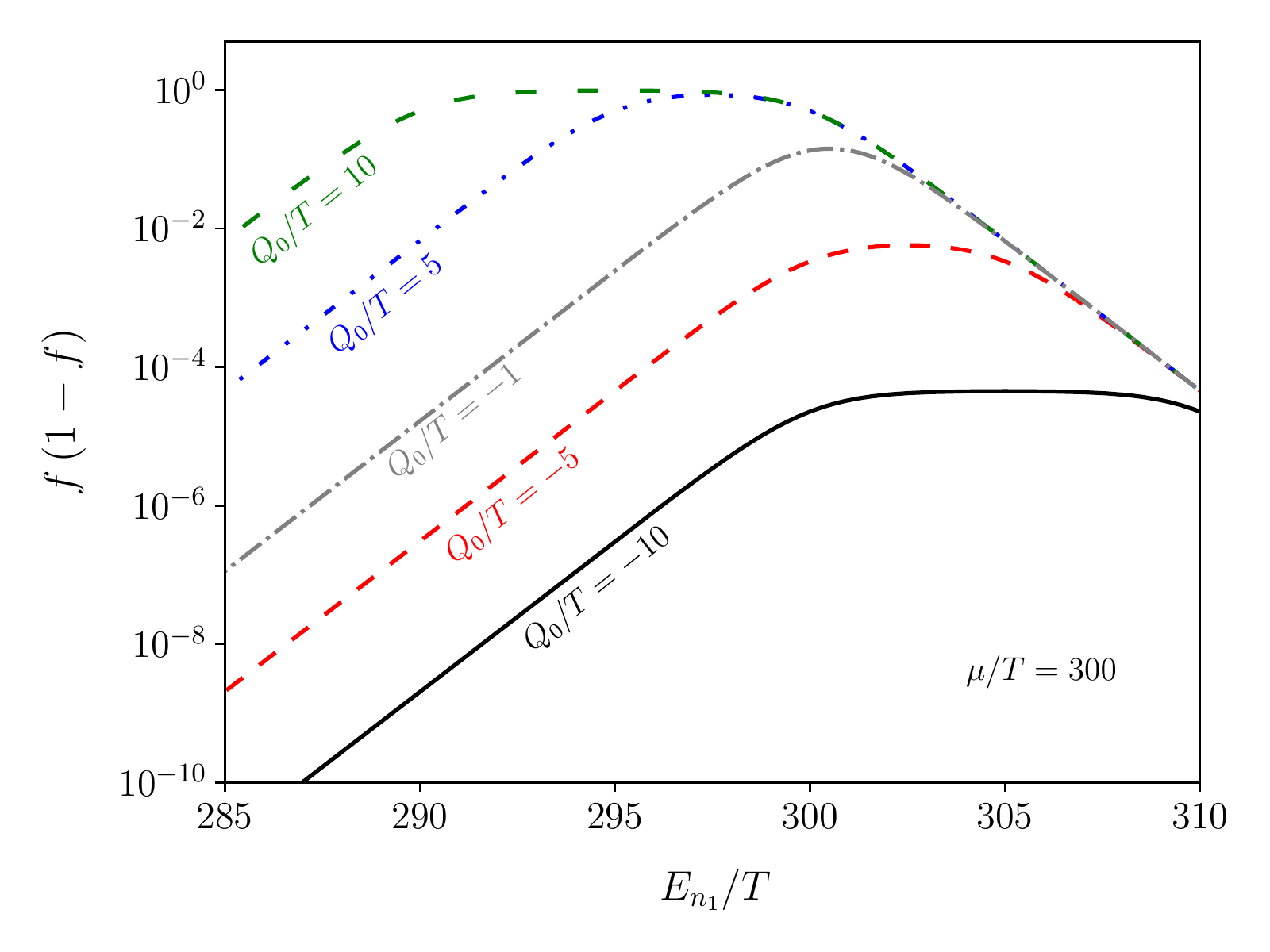}
 \caption{Plot demonstrating degeneracy suppression in a Fermi degenerate matter. We plot the factor that leads to this suppression namely $f(E_{N_1})\,(1-f(E_{N_1}+Q_0))$, where $E_{N_1}$ is the kinetic energy of the neutron in the initial state. We find it useful to plot with respect to dimensionless quantities. Here $f$ is the usual Fermi-Dirac distribution given by $f = 1/(e^{E_{N_1}/T - \mu/T}+1)$ and $T$ is the neutron star temperature. The area under the curve is suppressed for large negative values of $Q_0/T$ as expected from Eq.\eqref{eq:SFINAL}.}
 \label{fig:Q0}
 \end{figure}

Although the form of the response in Eq.\eqref{eq:mainS} is similar to those calculated previously, there are however, important differences. In the previous studies the response function was calculated for $2 \leftrightarrow 2$ processes and therefore the transfer energy $Q_0$ could in principle be as large as possible. But in our case, as will be discussed in Sec.\,\ref{appendix:emissivity}, its maximum value can be $\sim E_\gamma / 2.5$. This follows from the kinematics and the fact that $|
\Bar{q}| > 0$. It is worthwhile to note that although $Q_0$ and consequently $z$ can take positive as well as negative values, the response function $S(Q_0, q)$ is always positive and thus it is best to write it as
\bea
S(Q_0, q)=\dfrac{M_N^2\,T}{\pi q}\Bigg |\dfrac{z}{1-e^{-z}}\Bigg|\,\Theta(\mu - E_-)\;.
\label{eq:SFINAL}
\eea
It is evident from Eq.\eqref{eq:SFINAL} that the response function is highly suppressed for negative values of $z=Q_0/T$. To understand this degeneracy suppression better, we plot in Fig.~\ref{fig:Q0} the product $f(E_{N_1})\,(1-f(E_{N_1}+Q_0))$ appearing in Eq.\eqref{eq:Stot}. We find that the area under the curve shrinks rapidly as $Q_0/T$ takes large negative values. The area increases and saturates to a maximum for positive values $Q_0/T$. As long as we are interested in cooling the star and not heating it, we expect that $Q_0 = E_{N_2}-E_{N_1}$ should not take large positive values because otherwise, a major part of the photon's energy goes into heating the final state neutron. As we will see in the next section the maximum value of $Q_0$ is $\sim E_\gamma \sim m_\gamma \sim T$.

\subsection{Emissivity}
\label{appendix:emissivity}
In this section, we use the response function derived above to calculate the emissivity. Plugging Eq.\eqref{eq:SFINAL} in the expression for emissivity gives   
\bea 
Q^{2\to 3} = \frac{64\,n_F M_N^2  \kappa^2}{4\,\pi}\frac{g_\gamma}{3}\int \frac{d^3p_\gamma}{(2\pi)^3}\,
\frac{f_\gamma p_\gamma^2}{2E_\gamma}
\int \frac{  d^3p_1\,d^3p_{2}}{4\,(2\pi)^6}  \frac{|Q_0| ( E_{1}+E_{2})}{|\vec{p}_\gamma - \vec{p}_1- \vec{p}_2|}\, \frac{\Theta(\mu - E_-)}{|1-e^{-Q_0/T)}|}\;.
\label{eq:Q1}
\eea 
As evident from Eq.\eqref{eq:SFINAL}, $Q_0$ cannot have large negative values as this would lead to an exponentially suppressed response function. Furthermore, the theta function in the same equation also suggests that $Q_0$ cannot be larger than $\sim \mu$ in a highly degenerate system. For a NS, $\mu \sim \mathcal{O}(\rm 0.1)$ GeV. Hence, for our purpose, we can safely consider $|Q_0| \ll M_N$. For elastic processes, we further have $Q_0 \sim q$. Taking all these into account, Eq.\eqref{eq:E_minus_plus} becomes
\bea 
E^- \approx  \sqrt{ M_N^2 + \bigg(\frac{M_NQ_0}{\sqrt{q^2-Q_0^2}}\bigg)^2}- M_N\;.
\eea 
Now, we have $0 < E_- < \mu$, where the upper bound is imposed by the theta function in Eq.\eqref{eq:Q1}.
$E_- >0$ translates into an upper limit on $q$ and we have $q_{\rm max} = \sqrt{2\, Q_0 \, M_N}$. Similarly, $E_- < \mu$ implies a lower limit on $q$ namely $q_{\rm min}=(Q_0\,(1+\xi))/\sqrt{\xi(2+\xi)}$, with $\xi = \mu/M_N$. Note, while deriving this, we only retained leading order terms in $Q_0/M_N$, but no approximation was made with respect to $\xi$. Using $\mu \sim 300$ MeV and $M_N\sim 1$ GeV, we find $q_{\rm min} \sim 1.5\,|Q_0|$. This defines the boundaries of the variable $q$ which we will use in what follows.

Moving forward, to simplify calculations, let us first define an angle $\phi$ between $\vec{p}_\gamma -\vec{p}_{1}$ and $\vec{p}_{2}$. Recall that $q^2 = (\vec{p}_\gamma -\vec{p}_{1}- \vec{p}_{2})^2$. We thus have,
\begin{eqnarray}
q^2 &= (\vec{p}_\gamma -\vec{p}_{1}- \vec{p}_{2})^2 = (\vec{p}_\gamma -\vec{p}_{1})^2 +(\vec{p}_{2})^2 - 2  {p}_{2} |\vec{p}_\gamma -\vec{p}_{1}| \cos \phi\;, \nn \\
&\Longrightarrow q\,dq = -p_{2}|\vec{p}_\gamma -\vec{p}_{1}| \,d (\cos \phi)\;.
\end{eqnarray}
Similarly, reiterating that $Q_0 = E_\gamma - E_{1}- E_{2}$, we have $dQ_0 = -dE_2$. In terms of the new variables $\phi$ and $Q_0$, the emissivity thus becomes
\begin{align}
Q^{2\to 3} = \dfrac{16\,n_F M_N^2  \kappa^2}{\pi} \int \dfrac{d^3p_\gamma}{(2\pi)^3}
\dfrac{p_\gamma^2 f_\gamma}{2E_\gamma}
\int \dfrac{d^3p_{1} q \,dq \,dQ_0}{4\,(2\pi)^5 \,q}  \dfrac{ (E_\gamma- Q_0-E_{1}) (E_\gamma-Q_0)}{|\vec{p}_\gamma -\vec{p}_{1}|}\bigg|\dfrac{Q_0}{1-e^{-Q_0/T}}\bigg|\;.
\end{align}
From the definition of $q$ its kinematic boundaries would be
\begin{align} 
q &\in  \bigg[|E_2-  |\vec{p}_\gamma -\vec{p}_{1}|| , |E_2+  |\vec{p}_\gamma -\vec{p}_{1}||\bigg]\;,  \nn \\
&= \bigg[|E_\gamma - E_1 -Q_0-  |\vec{p}_\gamma -\vec{p}_{1}|| , |E_\gamma - E_1 -Q_0+  |\vec{p}_\gamma -\vec{p}_{1}||\bigg]\;. 
\end{align} 
Since the massive photon inside the NS is non-relativistic while the outgoing neutrinos are relativistic, we can take $|\vec p_\gamma| \ll |\vec p_1|$. The kinematic boundaries of $q$ thus simplify to
\begin{eqnarray}
q \in  \bigg[|E_1-E_2| , E_1 + E_2\bigg]\;.
\end{eqnarray} 
Taking into account the previous limits derived from the neutron response function, the final lower limit of $q$ becomes Max~$[1.5|Q_0|,|E_1-E_2|]$ while the upper limit is given by Min~$[|E_\gamma-Q_0|,\sqrt{2\,Q_0\,M_N]}$. The incoming photon is non-relativistic and in the NS rest frame our process is well approximated by simply its decay to neutrinos at rest \cite{Harvey:2007rd}. This implies $E_1 \simeq E_2$. 
Also, for the process under consideration, $|Q_0| \sim T$. Furthermore, as discussed in Sec.~\ref{section:SMproc}, in typical neutron stars, $m_\gamma \sim T$ and since the photons are non-relativistic and in equilibrium with the NS material, $E_\gamma \sim m_\gamma \sim T$. Hence the final range of integration of $q$ is given by
\bea 
q \in [1.5 |Q_0|,|E_\gamma  -Q_0| ]\;.
\eea 
It follows from the definition of $Q_0$ that its range of variation is $-\infty$ to $E_\gamma- E_{1}$. Let us hence split the $Q_0$ integral into two pieces 
\bea 
\int \limits_{-\infty}^{E_\gamma} (\ldots)\,dQ_0 =  \underbrace{ \int \limits_{0}^{E_\gamma} (\ldots)\,dQ_0}_{\text{positive } Q_0} + \underbrace{ \int \limits_{-\infty}^{0} (\ldots)\, dQ_0 }_{\text{negative } Q_0}\;, 
\eea 
The integral over $q$ is trivial, however, one has to be careful about the sign of $Q_0$ while performing it. We have the following two regimes,
 \begin{itemize}
     \item  \underline{Case 1}: $Q_0> 0$, 
\bea 
\int^{E_\gamma  -Q_0}_{1.5 Q_0} dq = (-1.5  Q_0 + E_\gamma  -Q_0) = (E_\gamma-2.5\, Q_0)\;.
\eea
\end{itemize}
We note that $\int dq$ is positive definite. This fixes the boundaries of $Q_0$. We hence find \,$0 \leq Q_0 \leq E_\gamma/2.5$. Similarly,
\begin{itemize}
\item \underline{Case 2}: $Q_0< 0$, the integral becomes 
\bea 
\int^{E_\gamma  -Q_0}_{1.5 |Q_0|} dq = (-1.5  |Q_0| + E_\gamma  -Q_0) = (E_\gamma-0.5 |Q_0|)\;,
\eea
\end{itemize}
and the range of $Q_0$ is now given by $-2\,E_\gamma \leq Q_0 \leq 0$. To summarise,
\bea 
\int dq = \begin{cases}
 (E_\gamma-0.5 |Q_0|) \qquad \text{for} \qquad Q_0 \in [- 2E_\gamma, \rm Min[0,E_\gamma -E_1]]\;,
    \\
    (E_\gamma-2.5 |Q_0|) \qquad \text{for} \qquad Q_0 \in [ 0, E_\gamma/2.5]\;.
    \label{eq:range_kin}
\end{cases}
\eea 
The full integral over $Q_0$ then becomes
\begin{align}
\mathcal{I}&= \int\limits^{0}_{-2 E_\gamma} \frac{d^3p_{1}  dQ_0}{4(2\pi)^5}  \frac{ (E_\gamma- Q_0-E_{1}) (E_\gamma-Q_0) (E_\gamma-0.5 |Q_0|)}{E_1}\bigg|\frac{Q_0}{1-e^{-Q_0/T}}\bigg| \nn
\\
&+ \int\limits_{0}^{E_\gamma/2.5} \frac{d^3p_{1}  dQ_0}{4(2\pi)^5}  \frac{ (E_\gamma- Q_0-E_{1}) (E_\gamma-Q_0)(E_\gamma-2.5 |Q_0|)}{E_1}\bigg|\frac{Q_0}{1-e^{-Q_0/T}}\bigg|\;. 
\end{align}
The integral contributes to the emissivity dominantly when $|Q_0| \lesssim T$. Hence, we expand the exponential to have $\bigg|\frac{Q_0}{1-e^{-Q_0/T}}\bigg| \approx T$. Performing the $Q_0$ integral along with $0 \leq E_1 \lesssim E_\gamma$ we finally get,
\begin{align}
 \mathcal{I}&=\frac{T}{32\pi^4}\int_{0}^{E_\gamma} E_1 dE_1 (3 E_\gamma^4 - 1.66 E_\gamma^3 E_1)
+ \frac{T}{32\pi^4}\int_{0}^{E_\gamma} E_1 dE_1  (-0.173 E_1 E_\gamma^3 + 0.152 E_\gamma^4) \nn \\
&\approx   \frac{1}{32\pi^4} E_\gamma^6 \, T + \frac{0.02}{32\pi^4} T\,E_\gamma^6 \approx  \frac{1}{32\pi^4} E_\gamma^6 \,T\;. 
\end{align}
Finally, performing the integral over $p_\gamma$ from 0 to $\infty$ and using $f_\gamma = 1/(e^{E_\gamma/T}-1) \sim e^{-E_\gamma/T} $, we get the emissivity of the neutron star due to the process under consideration as
\begin{eqnarray}
Q^{2\to 3}  &=& \dfrac{48\,n_F\, M_N^2  \,\kappa^2}{3(2\,\pi)^5}\dfrac{ 3 m_\gamma^5 T^3}{(2\pi)^2 }\bigg( (m_\gamma^3 + 33 m_\gamma T^2)K_4(m_\gamma/T)+ 9T(m_\gamma^2 + 35 T^2)K_5(m_\gamma/T)\bigg) \nn \\
&=&\dfrac{48\,n_F \, M_N^2 \, \kappa^2}{(2\,\pi)^7}\,T^{11}\,\underbrace{x^5\,\left( (x^3 + 33x)K_4(x)+ 9(x^2 + 35 )K_5(x)\right)}_{\equiv \mathcal{H}(x)}\;,
\end{eqnarray}
where, $x \equiv m_\gamma/T$ and $K_4 (x)$ and $K_5(x)$ are modified Bessel functions of order 4 and 5 respectively. The function $\mathcal{H}(x)$ is $\sim \mathcal{O}(10^5)$ for $x \lesssim 5$. For $x \gg 5$, $\mathcal{H}(x) \sim e^{-x}\,\sqrt{\pi/2}\,x^{15/2} $
The limiting behaviour of the emissivity can thus be summarised as
\begin{equation}
  Q^{2\to 3} =
    \begin{cases}
      1.2\times 10^{5} \,\dfrac{48\,n_F \, M_N^2 \, \kappa^2}{(2\,\pi)^7}\,T^{11} & \text{for $x \lesssim 5$}\;,\\
      \dfrac{48\,n_F \, M_N^2 \, \kappa^2}{(2\,\pi)^7}\,\sqrt{\dfrac{\pi}{2}}\,T^{7/2}\,m_\gamma^{15/2}\,e^{-m_\gamma/T}& \text{for $x \gg 5$}\;.\\
    \end{cases}  
    \label{eq:Qfinal}
\end{equation}

\begin{figure}[h]
\centering
 \includegraphics[scale=0.7]{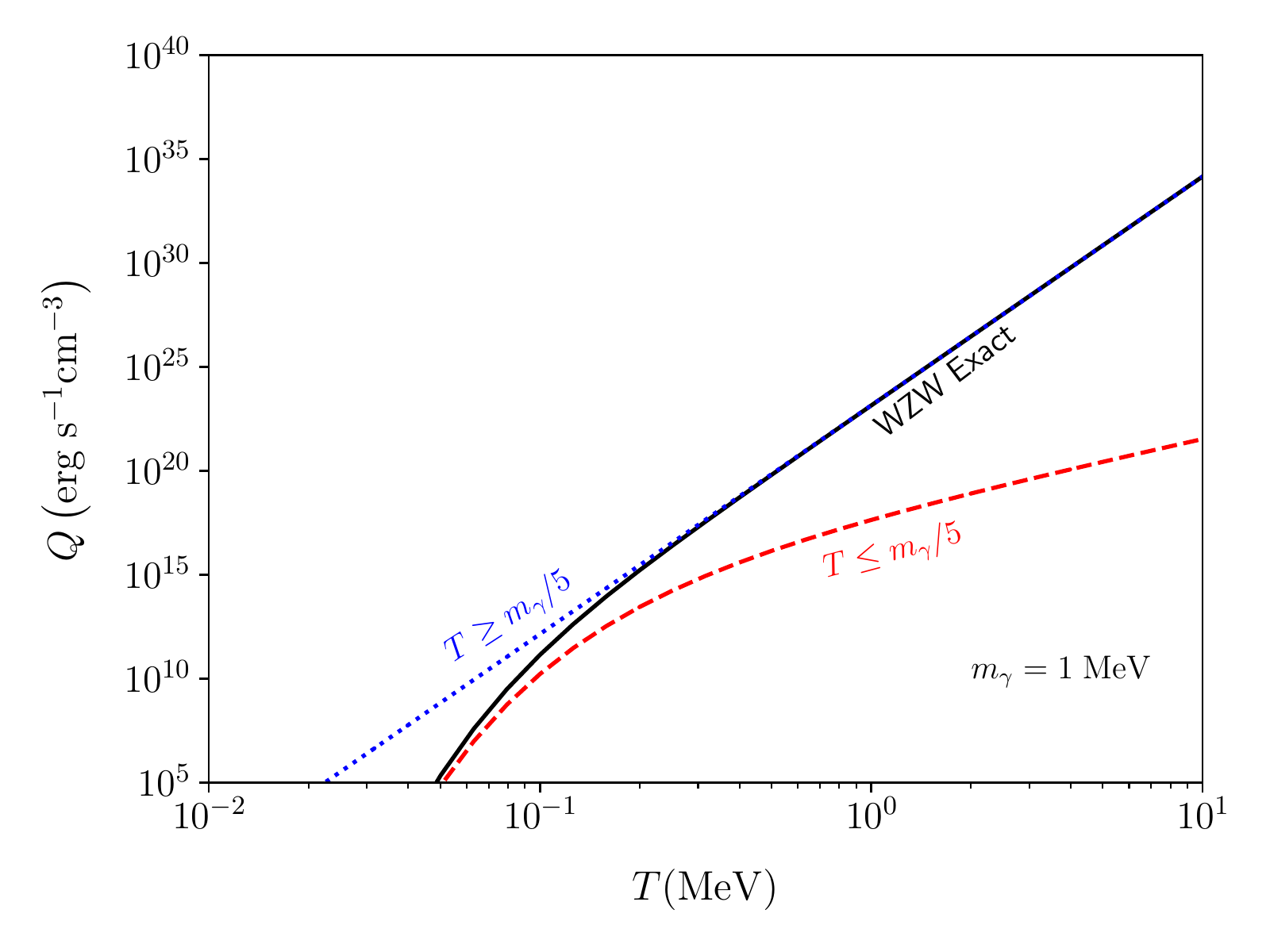}
 \caption{The emissivity for the cooling process with a 1 MeV photon in the initial state is plotted as a function of the temperature of the neutron star. The limiting behaviours are shown with dashed lines and they are found to agree well with the actual result in suitable regimes.}
 \label{fig:emissivity}
 \end{figure}
These are plotted in Fig.~\ref{fig:emissivity} for completeness. 
The $T^{11}$ dependence of the emissivity as seen in Eq.\eqref{eq:Qfinal} can be understood as follows: The response function depends only on the background neutrons and their level of degeneracy. In the elastic regime $|Q_0| \sim q \sim T$ and hence $S \propto M_N^2$. The Lorentz invariant phase space measure $d^3 p/(2E)$ of light particles, i.e., when $T \gtrsim$ their respective masses, varies as $T^2$. At large $T$, the neutrino energies $E_1,\,E_2$ and the photon momentum $p_\gamma$ all scale as $T$. Therefore, the scattering matrix in Eq.\eqref{eq:Msqr} at large temperatures goes as $T^4$. Finally, the emissivity as given in Eq.\eqref{eq:Def:cooling_rate} is obtained by multiplying the scattering matrix with the total energy that is being transported out of the system, i.e., $E_1+E_2 \sim T$. Taking all the $T$ dependence into account we thus have 
\bea 
Q \propto \underbrace{T^4}_{|\mathcal{M}|^2} \times \underbrace{T^{2}}_{\text{First $\nu$}}  \times\underbrace{T^{2}}_{\text{Second $\nu$}} \times \underbrace{T^{2}}_{\text{$\gamma$}}\times \underbrace{T}_{\text{Energy loss}} \propto T^{11} \, . 
\eea 

\subsection{Results and Discussion}

The cooling induced by a massive photon in the background of degenerate neutrons was estimated in \cite{Harvey:2007rd} using a zero-momentum exchange approximation. Treating the neutron as infinitely massive fixed targets, the scattering was replaced by a simple decay of the photon $\gamma \rightarrow \nu \Bar{\nu}$. 
\bea 
Q^{\text{\cite{Harvey:2007rd}}} = 3 \int \frac{d^3p}{(2\pi)^3} E_\gamma f_\gamma(E_\gamma) \Gamma_{\gamma \to \nu \nu}\;, \; \; \; \text{with} \; \; \; \Gamma_{\gamma \to \nu \nu} = \frac{2 m_\gamma \kappa^2 n_B^2}{9\pi}p_\gamma^2\;.
\label{eq:QHarvey0}
\eea 

Using $E_\gamma \approx m_\gamma + p_\gamma^2/2m_\gamma$ the authors of \cite{Harvey:2007rd} found the emissivity to be 
\bea 
\label{eq:QHarvey1}
 Q_{m_\gamma \gg T}^{\text{\cite{Harvey:2007rd}}} = 
     \frac{\sqrt{2\pi} \alpha g_\omega^4}{16\pi^6} \frac{G_F^2 m_\gamma^2 n_B^2}{m_\omega^4} e^{-m_\gamma/T} (m_\gamma T)^{5/2} \qquad m_\gamma > T  \,. 
\eea
We generalise this formula to include regimes where $m_\gamma < T$. We find
\bea
\label{eq:harveyHT}
Q^{\text{\cite{Harvey:2007rd}}}=\dfrac{T^3}{\pi^3}m_\gamma^4 \kappa^2 n_B^2 \left(\dfrac{m_\gamma}{T}K_2 (m_\gamma/T)+5K_3 (m_\gamma/T)\right) \, . 
\eea
In the limit when $m_\gamma \ll T$ this expression simplifies to 
\bea
Q = \dfrac{40 m_\gamma \kappa^2 n_B^2}{\pi^3}\,T^6=\dfrac{5\,m_\gamma T^6}{\pi^6}\dfrac{g_\omega^4\alpha G_F^2}{m_\omega^4} \,. 
\eea

The decay takes place mostly in the regions of the neutron star with density $\sim 2\rho_0$\,\cite{Harvey:2007rd}. However, the by-standing neutrons in the process at such densities are mostly degenerate and the effect of degeneracy suppression due to Pauli blocking should be included in the calculation in a systematic way. In the rest of this section, we try to include the suppression as a simple multiplicative factor with the result obtained in Eq.\eqref{eq:QHarvey1} and compare the results with the actual calculation carried out in this paper. The degeneracy suppression factor as defined in \cite{Raffelt:1996wa} is given by
\bea 
F_{\rm Deg} = \dfrac{2}{n_B}\int d^3 p \,f_{\rm FD}(p) \,(1-f_{\rm FD}(p)) \simeq  3\,\frac{E_F\,T}{p_F^2} \simeq \frac{3\,T}{2\,M_N}\;,
\eea 

where we used that $E_F \simeq p_F^2/2M_N$ and $n_B \simeq p_F^3/3\pi^2$. Including this in Eq.\eqref{eq:QHarvey1} we obtain
\bea 
 Q^{\text{\cite{Harvey:2007rd}+Deg} } = \dfrac{3T^4}{2\pi^3}m_\gamma^4 \kappa^2 \frac{n_B^2}{M_N} \left(\dfrac{m_\gamma}{T}K_2 (m_\gamma/T)+5K_3 (m_\gamma/T)\right)\;. 
\label{eq:QHarveyCorr}
\eea 

\begin{figure}[!h]
\centering
 \includegraphics[scale=0.7]{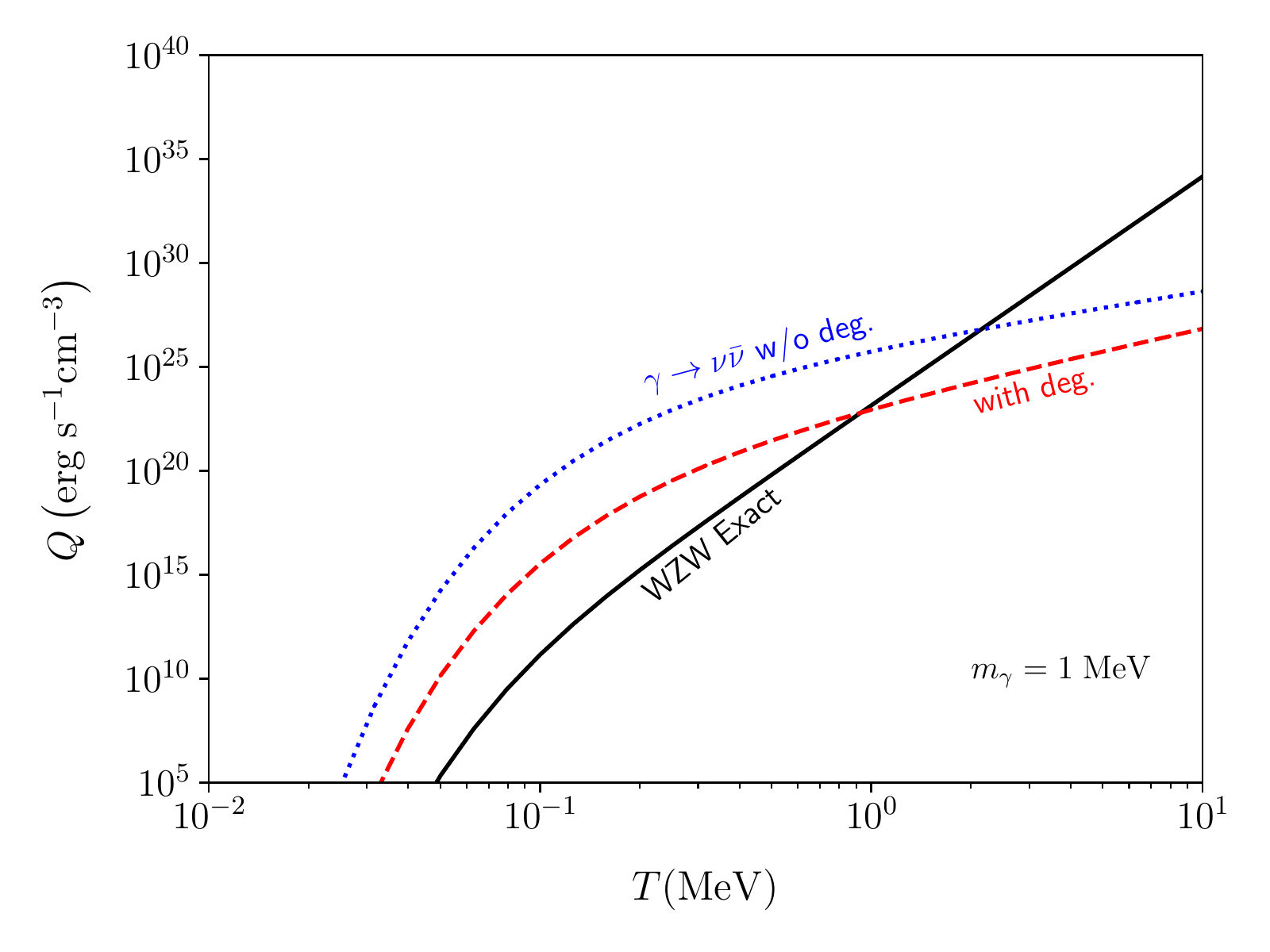}
 \caption{Comparison of the computation presented in this paper with earlier computations for a 1 MeV photon. The blue curve is the emissivity plotted using the corresponding expression in \cite{Harvey:2007rd}. The red line is the same multiplied by an overall degeneracy suppression factor of $\sim T/m_N$. The black line represents the emissivity that has been calculated in this work.}
 \label{fig:compare}
 \end{figure}

\begin{figure}[!h]
\hspace{-0.9cm}
 \includegraphics[scale=0.5]{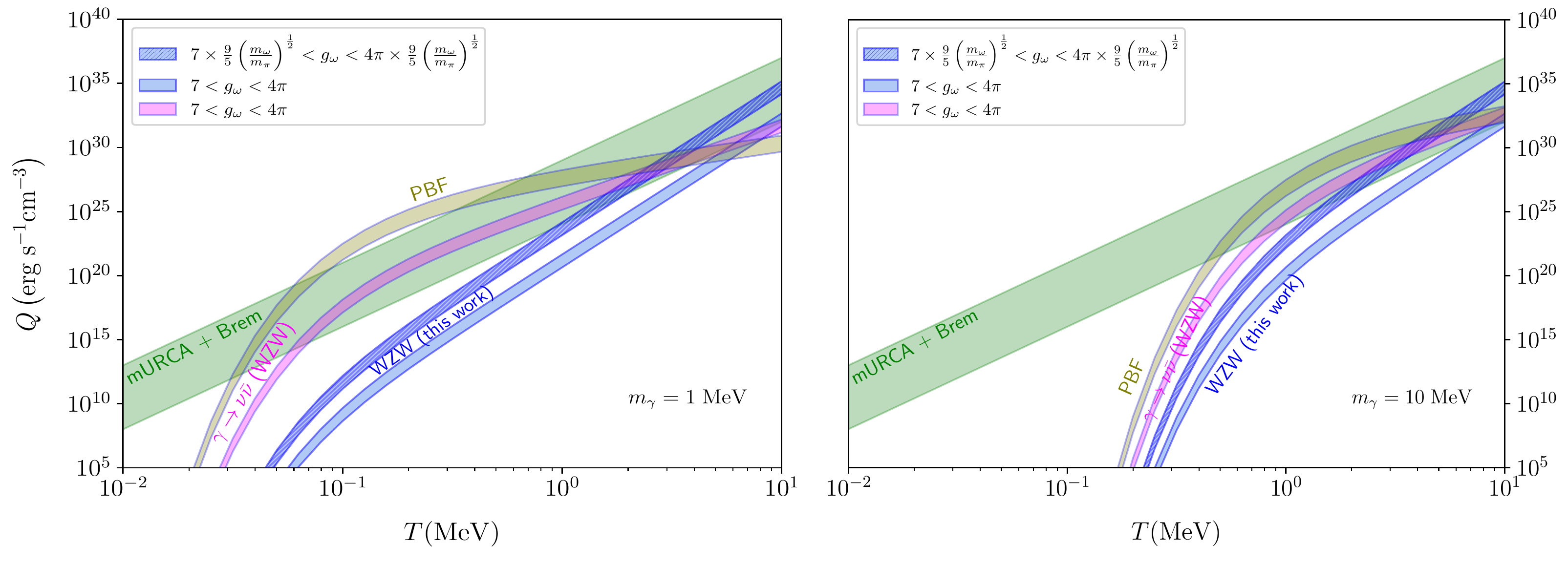}
 \caption{Comparison of emissivities due to the WZW term with other conventional processes like mURCA~\cite{Yakovlev:2000jp}, PBF~\cite{Voskresensky_1998}, $\gamma\to\nu\bar\nu$ (WZW) (Eq.~\ref{eq:harveyHT}) and $\nu$ bremsstrahlung~\cite{Yakovlev:2000jp,1979ApJ...232..541F,1995A&A...297..717Y} for a 1 MeV photon (left) and for a 10 MeV photon (right). The bands depict the theoretical uncertainties involved in such processes. For mURCA and $\nu$ bremsstrahlung, the uncertainties mostly arise from our lack of knowledge about the medium effects. For WZW, we give different values of the $g_\omega$. Finally, for PBF, the density is varied between $\rho_0$ and $3\rho_0$. Further note that since $T_c \sim 1$ MeV for nucleon pair formation, the shaded region for PBF in principle should not extend much beyond 1 MeV.}
 \label{fig:uncertn2}
\end{figure}

We plot the previous result with (Eq.\eqref{eq:QHarveyCorr}) and without (Eq.\eqref{eq:QHarvey1}) the naive inclusion of degeneracy and compare it with the actual computation. All these are shown in Fig.~\ref{fig:compare}. We see, as expected, the simplistic approach to include degeneracy as an overall factor does not agree well with the actual calculation denoted by the blue line. Furthermore, we find that the result of \cite{Harvey:2007rd} indicated by the red dashed line in the plot is an overestimation due to the non-inclusion of the degeneracy suppression for temperatures~$\lesssim$ MeV. From the discussions above we expect that the emissivity is not degeneracy suppressed for $T \gg m_N$. This implies that our result should be asymptotically similar to the result obtained in \cite{Harvey:2007rd} since the latter does not take the effect of degeneracy into account. However, from Fig.~\ref{fig:compare} we find that this is not the case. In the high temperature limit our calculation of emissivity shows a $T^{11}$ dependence while that of \cite{Harvey:2007rd} goes as $T^6$. This discrepancy is expected since in our computation we allowed for a finite exchange of momentum between the nucleons and the $\gamma \to \nu\nu$ reaction, which was forbidden by construction in \cite{Harvey:2007rd}.
In plotting our result as well as those using Eq.\eqref{eq:QHarvey1} we presented different values of the coupling $g_\omega $. To have an idea of how well the WZW term competes with the mURCA and the $\nu$-bremsstrahlung within these uncertainties, we plot Fig.~\ref{fig:uncertn2}. The uncertainties may a priori seem large, however, thanks to the strong temperature dependence of the emissivities, these can easily be counterbalanced by a small change of the same. We find that the contribution from the WZW term can compete with the $\nu$-bremsstrahlung for $T \gtrsim 0.5$ MeV. On the other hand, cooling via WZW becomes comparable to mURCA at $T \sim$ MeV. These statements largely remain true even if we vary the mass of the photon. We show this by presenting a similar plot with a $10$ MeV photon in Fig.~\ref{fig:uncertn2}. At this point, we remind ourselves that our calculation uses the fact that the photons are non-relativistic and almost at rest with respect to the neutrons and NS rest frame. This implies $T$ should not be too large than $m_\gamma$. Therefore, we do not extend the $x$-axis much beyond $10$ MeV, a typical photon mass as seen from the discussions in Sec.~\ref{section:SMproc}. Also, for neutron stars which are much hotter, the neutrinos cannot stream out freely. Cooling under such circumstances will take place via scatterings and absorption of neutrinos.

\section{Conclusion and future directions}
\label{sec:conclusion}
The WZW term is unavoidable within the realms of the Standard Model. In this work, we have discussed the effect of such an interaction in the cooling of a hot neutron star. Specifically, we have shown that due to the $\epsilon^{\mu \nu \alpha \beta} \, F_{\mu \nu} \,\omega_\alpha \, Z_\beta$ term in the Lagrangian neutrinos can be emitted from the core of a neutron star via $N \,\gamma \to N\, \nu \,\Bar{\nu}$. We have calculated the emissivity of such a process keeping in mind the suppression due to the highly degenerate neutrons in the initial and final state. We find that this can contribute significantly to the cooling of the NS for $T \gtrsim 1$ MeV compared to other standard processes like mURCA, PBF and $\nu$ bremsstrahlung. We also found that the inclusion of degeneracy by a simplistic multiplicative factor does not correctly reproduce the actual calculation. As compared to the simplistic treatment of \cite{Harvey:2007rd} which inferred that such cooling mechanisms are less likely to be important even for cold neutron stars, our detailed calculation suggests that it can be a dominant cooling channel only for very young and hot neutron stars and if the coupling $g_\omega$ is very large $\gtrsim 20$. In passing, we also note that contrary to the plasmon decay~\cite{Harvey:2007rd} and PBF~\cite{Voskresensky_1998}, which would vanish in the limit of zero photon mass, the $2\to 3$ process under study is kinematically allowed even for small photon masses. At this point one might think that the plasma effects discussed in \cite{1967ApJ...150..979B} in the context of photo-neutrino emission would provide a further suppression to the result we presented here. However as discussed in \cite{1967ApJ...150..979B}, this suppression is important in the regime $\omega_P/T \sim m_\gamma/T \gg 1$ and becomes negligible in the opposite regime, i.e., $\omega_P/T \sim m_\gamma/T \ll 1$. It is in this latter regime that the emission due to the WZW type of interaction as discussed here becomes most relevant and competes with standard cooling processes. Therefore in this regime, we can safely neglect any suppression via plasma corrections.

The effect of WZW interactions on the equation of state of a NS would be interesting to study and will be investigated in a future publication. Moreover, In~\cite{Hannestad:1997gc} and \cite{Li:2020ujl}, authors have shown that taking into account flavour-blind interactions like neutron bremsstrahlung might have a strong impact on the relative ratio of $\nu_\tau, \nu_\mu, \nu_e$ emitted. Noticing that the WZW term is also flavour blind, we might expect that this type of interaction might as well modify the relative ratio at high temperatures. 

In the context BSM physics, the presence of WZW interactions might also open some new channels of cooling. For example, it has been shown in \cite{DiLuzio:2017ogq, Badziak:2023fsc, Takahashi:2023vhv} that axion can be made naturally decoupled from nucleons, making them effectively {\it astrophobic} and allowing to lower the bound on the decay constant $f_a$ from the cooling argument as low as $f_a \gtrsim 10^{6}$ GeV. It should be investigated if the unavoidable presence of the WZW term would lead to reconsidering this conclusion~\cite{Chakraborty:2024tyx}. 

\section*{Acknowledgements}

AG is supported by the ``Generalitat Valenciana" through the GenT Excellence Program (CIDEIG/2022/22). MV is supported by the ``Excellence of Science - EOS" - be.h project n.30820817 and by the Strategic Research Program High-Energy Physics of the Vrije Universiteit Brussel.  We would also like to thank Nicolas Chamel, Arijit Kundu, Nilay Kundu, Alberto Mariotti and Sanjay Reddy for their insightful discussions. SC would like to acknowledge the hospitality of IACS, Kolkata where the final part of the project was completed.

\appendix

\section{Computation of the response function}
\label{app:computation}

In this appendix, we compute the response function $S$. By definition, we have 
\begin{align}
S(q^\mu) &= g_N \int \dfrac{d^3p_{N_1}d^3p_{N_2} M_N^2}{(2\pi)^6\, E_{N_1}\,E_{N_2}} f_N(E_{N_1})(1-f_N(E_{N_2})) \times (2\pi)^4 \times \nonumber \\
&\delta(E_{N1}+Q_0-E_{N_2})\,\delta^3(\vec{p}_{N_1}-\vec{p}_{N_2}- \vec{q}) \,\Theta(E_{N_1}-M_N) \,\Theta(E_2-M_N)\;, \nonumber \\
&= \int \frac{2\pi p\, E_{N1}\,dE_{N1} \,d(\cos \theta) \,M_N^2}{2\pi^2 E_{N1}\,\sqrt{E_{N1}^2 + q^2 + 2 q\,|\vec{p}_{N_1}|\cos \theta}} f_N(E_{N1})\,(1-f_N(E_{N1}+Q_0))\nonumber \\
&\times \delta(E_{N1}+Q_0-E_{N_2}) \,\Theta(E-M_N)\, \Theta(E+Q_0-M_N)\;,
\label{S1}
\end{align}
where, in the last step, we have performed the integral over $\vec{p}_{N_2}$. We have used that $ \vec{p}_{N_2} = \vec{p}_{N_1}+\vec{q}$ and $E_{N_2} = E_{N_1}+ Q_0 = \sqrt{M_N^2 +(\vec{p}_{N_1}+\vec{q})^2} = \sqrt{E_{N_1}^2 + q^2 + 2 q\,|\vec{p}_{N_1}| \cos \theta}$ with $\theta$ being the angle between $\vec{p}_{N_1}$ and $\vec{q}$.
Next, we perform the $\cos \theta$ integral. 
We have
\bea
\delta(E_{N1}+Q_0-E_{N_2})= \dfrac{E_{N_2}}{|\vec{p}_{N_1}|\,q}\delta(\cos\theta-\cos\theta_0)\;,
\eea
where,
\bea 
\cos \theta_0 = \frac{Q_0^2-q^2+2EQ_0}{2q\sqrt{E^2-M_N^2}}\;.
\label{eq:angle}
\eea 
Using these, the response function in Eq.~(\ref{S1}) simplifies to 
\bea 
S(Q_0, q)=\frac{  M_N^2}{\pi  q}\int_0^{\infty} dE f_N(E)(1-f_N(E+Q_0))  \Theta(E-M_N) \Theta(E_{N_2}-M_N)\;.
\eea 
To integrate over $E$, we first need to fix the limit of the integration. The cosine of the angle in Eq.\eqref{eq:angle} lies between -1 and 1 and this in turn provides the required range of integration. We hence have,
\begin{align}
& S(Q_0, q) = \frac{  M_N^2}{\pi  q}\bigg[\int_{E^-}^{\infty} dE f_N(E)(1-f_N(E+Q_0))  \Theta(E-M_N) \Theta(E_{N_2}-M_N)\Theta(q^2 - Q_0^2)
\nn \\
&+ \int_{0}^{E^+} dE f_N(E)(1-f_N(E+Q_0))  \Theta(E-M_N) \Theta(E_{N_2}-M_N)\Theta( Q_0^2- q^2) \bigg]\;,
\label{eq:Stot}
\end{align}
with,
\bea 
E^- = \sqrt{\bigg(M_N+\frac{Q_0}{2}\bigg)^2 + \bigg(\frac{\sqrt{q^2-Q_0^2}}{2}-\frac{M_NQ_0}{\sqrt{q^2-Q_0^2}}\bigg)^2}- \bigg(M_N+\frac{Q_0}{2}\bigg)\;,
\eea
and
\bea
E^+ = \sqrt{\bigg(M_N+\frac{Q_0}{2}\bigg)^2 - \bigg(\frac{\sqrt{q^2-Q_0^2}}{2}+\frac{M_NQ_0}{\sqrt{q^2-Q_0^2}}\bigg)^2} - \bigg(M_N+\frac{Q_0}{2}\bigg)\;.
\label{eq:E_minus_plus}
\eea 
The second line of Eq.(\ref{eq:Stot}) contributes only when the exchange of energy ($Q_0$) and momentum ($|\vec{q}|$) is of the order of the neutron mass or more, i.e, in the inelastic regime. For our purpose $Q_0 \sim |\vec{q}| \ll M_N$ and hence this contribution can be safely neglected. Thus, using the Fermi-Dirac distribution for the neutrons and performing the integral in Eq.(\ref{eq:Stot}) over E, we get
\bea
S(Q_0, q)=\dfrac{M_N^2\,Q_0}{\pi q}\dfrac{e^{Q_0/T}}{e^{Q_0/T}-1}\left[\dfrac{T}{Q_0}\ln\left(\dfrac{1+e^{(E_- - \mu)/T}}{1+e^{(E_- +Q_0 - \mu)/T}}\right)+1\right]\;.
\label{eq:FinaS}
\eea
where $T$ is the neutron star temperature. In highly degenerate matter, the term in the parenthesis can simply be replaced by $\Theta(\mu - E_-)$~\cite{Reddy:1997yr}. Defining $z = Q_0/T$, we get the final form of the response function
\bea
S(Q_0, q)=\dfrac{M_N^2\,Q_0}{\pi q}\dfrac{1}{1-e^{-z}}\Theta(\mu - E_-)\;.
\label{eq:Final2S}
\eea

\section{Wess-Zumino-Witten Interactions}
\label{app:WZWterm}
Chiral Lagrangian ``effectively'' describes the dynamics of the Goldstone bosons associated with the spontaneous breaking of the symmetries of QCD. At the leading order, it contains the kinetic terms, exhibiting the same chiral symmetry as QCD. In addition, symmetry-breaking effects are included as spurions. Together, the Lagrangian has the form:
\begin{equation}
    \mathcal L = \frac{f_\pi^2}{4}\text{Tr}\left(D_\mu U^\dagger D^\mu U+U^\dagger\chi+\chi^\dagger U\right)\;,
\end{equation}
where
\begin{align}
    D_\mu = \partial_\mu U - i r_\mu U + i U \ell_\mu\;, \; \;  \; r_\mu(\ell_\mu) = v_\mu \pm a_\mu\;, \; \; \; \chi = 2 B_0 \mathcal M\;.
\end{align}
Here, $\mathcal M$ denotes the quark mass matrix. $B_0$ and $f_\pi$ are constants that are fixed by experiments and not by any symmetry requirements. As it happens, this effective theory might have more symmetries as compared to the UV theory, i.e., QCD. Therefore, it misses several important terms including anomalies. This was first pointed out by Wess and Zumino~\cite{Wess:1971yu} and later represented in a geometrical way by Witten~\cite{Witten:1983tw}. It turns out, such a term encapsulating the anomalies can not be written in four dimensions. Instead, we have to consider a 5-dimensional action where the boundary would be identified with our 4-dimensional space, i.e.,
\begin{equation}
    S_{\text{WZW}} = \kappa \int_D d^5 y\; \omega\;,
\end{equation}
where
\begin{equation}
    \omega = -\frac{i}{240\pi^2} \epsilon^{\mu\nu\rho\sigma\tau}\; \text{Tr}\left(U^\dagger \partial_\mu U\; U^\dagger \partial_\nu U\; U^\dagger \partial_\rho U\; U^\dagger \partial_\sigma U\; U^\dagger \partial_\tau U\right)\;.
    \label{eq:WZW_Action}
\end{equation}
Here, in terms of pion fields, $U=\text{exp}\left(2i \pi^a T^a/f_\pi\right)$\; and transforms linearly under the chiral symmetry $SU(3)_L\times SU(3)_R$ as $U\to R U L^\dagger$. Expanding the WZW action leads to interactions between five Goldstone particles, leading to processes such as $K^+K^-\to\pi^+\pi^-\pi^0$.

\subsection{Gauging the $U(1)_{\text{em}}$ subgroup} The straightforward approach of defining the covariant derivative fails for WZW action. The principal reason being all the gauging must be done in the 4-dimension, i.e., at the boundary of the 5-dimensional ball. Therefore, one resort to the {\it trial and error} approach by noticing the change of WZW action under infinitesimal transformation 
\begin{equation}
    \delta U = i \epsilon(x)\left[Q, U\right]\;.
    \label{eq:trans}
\end{equation}
The parameter $\epsilon(x)$ depends only on the 4-dimensional coordinates and $Q$ is the $SU(3)$ charge matrix. Under the transformation shown in Eq.~(\ref{eq:trans}), the WZW action changes as
\begin{equation}
    \delta\omega = \partial_\mu \epsilon(x) \hat{J}^\mu\; = e\; \delta{A_\mu} \hat{J}^\mu\;,
    \label{eq:gauge_variation}
    \end{equation}
where, 
\begin{equation}
\hat J^\mu = \frac{1}{48\pi^2}\; \epsilon^{\mu\nu\rho\sigma\tau}\; \partial_\nu\; \text{Tr}\left[\{Q,U^\dagger\}\;\partial_\rho U\; U^\dagger\partial_\sigma U\; U^\dagger \partial_\tau U\right]\;.
\label{eq:variation}
\end{equation}
Since the current given in Eq.~(\ref{eq:variation}) is a total derivative, the variation of the WZW term reduces to a boundary term which can be cancelled by the variation of the 4-dimensional gauge field. In other words, it is equivalent to adding $\int d^4 x A_\mu J^\mu$ term~\footnote{Note that $J^\mu$ is the 4-dimensional current.} in the Lagrangian, whose variation should cancel Eq.~(\ref{eq:gauge_variation}). However, it is rather obvious from the form of the current that $\delta J^\mu\neq 0$. Hence, one needs to include compensating terms to make the whole WZW action gauge invariant. Finally, we obtain
\begin{align}
S_{\text{WZW}}\left(U,A_\mu\right) = & \kappa \Bigl[\int_D d^5 x\; \omega - e\int d^4 x\; A_\mu J^\mu \\
& + \frac{i e^2}{24\pi^2}\int d^4 x\;  \epsilon^{\mu\rho\sigma\lambda}\; A_\rho \left(\partial_\mu A_\nu\right)\; \text{Tr}\left(\{Q^2,U^\dagger\}\partial_\sigma U\right)-Q U Q \partial_\sigma U^\dagger\Bigr]\;.
\label{eq:gaugingU1}
\end{align}
The action in Eq.~(\ref{eq:gaugingU1}) reproduces the anomaly term $\pi^0 F_{\mu\nu} \tilde{F}^{\mu\nu}$ when one identifies the coefficient $\kappa$ with the color factor. As far as applications to the Standard Model are concerned, gauging subgroups other than $U(1)_{\text{em}}$ becomes important and we will discuss this in the next section.

\subsection{Gauging non-abelian subgroup}
We follow the same {\it trial and error} method to gauge arbitrary subgroup of $SU(3)_L\times SU(3)_R$. The transformation properties are tabulated below
\begin{equation}
    \delta U = i\left(\epsilon_L U - U \epsilon_R\right)\;, \; \; \; \delta A_{(L,R)}^a = - \frac{1}{g} \partial_\mu \epsilon^a_{(L,R)} + f^{abc} \epsilon_{(L,R)}^b A_\mu^c\;.
\end{equation}
For notational simplification, we use $\epsilon = \epsilon^a T^a$, where $T^a$'s are the corresponding generators. Just like in the previous case, the variation of the WZW action defined in Eq.~(\ref{eq:WZW_Action}) gives
\begin{equation}
    \delta \omega = \frac{1}{48\pi^2}\; \epsilon^{\mu\nu\rho\sigma\tau}\; \text{Tr}\left[\partial_\mu\epsilon_L^a\; \hat{J}_L^{\mu a}+\left(L\to R\right)\right]\;,
\end{equation}
where
\begin{eqnarray}
    \hat{J}_L^{\mu a} = \frac{1}{48\pi^2}\; \epsilon^{\mu\nu\rho\sigma\tau}\;\partial_\nu\;\text{Tr}\left[T_L^a\; U_{\rho_L}U_{\sigma_L}U_{\tau_L}\right]\;, \nonumber \\
    \hat{J}_R^{\mu a} = \frac{1}{48\pi^2}\; \epsilon^{\mu\nu\rho\sigma\tau}\;\partial_\nu\;\text{Tr}\left[T_R^a\; U_{\rho_R}U_{\sigma_R}U_{\tau_R}\right]\;. \nonumber \\
\end{eqnarray}
Here, we use the shorthand notation for $U_{\mu_L} = (\partial_\mu U)U^\dagger$ and $U_{\mu_R}=U^\dagger (\partial_\mu U)$\;. Again, following the previous calculation, the variation of this current does not vanish which in turn requires appropriate compensating terms. The complete result, given in terms of the action $S_{\text{WZW}}(U,A_L,A_R)$ is tabulated in a convenient form in~\cite{Witten:1983tw, Kaymakcalan:1983qq, Chou:1983qy, Kawai:1984mx, Pak:1984bn}. As mentioned before, we are interested in interactions between the fundamental gauge fields ($\gamma, Z$) with background fields ($\omega$), which can be included in the effective action by the transformation: $\mathcal A_{L,R}=A_{L,R}+B_{L,R}$, where $A, B$ are the fundamental and background gauge fields respectively~\cite{Harvey:2007ca}. From the full set of terms, the interactions relevant to our process come from
\begin{align}
    S_{\text{WZW}} \left(U,\mathcal A_{L,R}\right) &\supset \frac{N_C}{48\pi^2} \int d^4 x\; \epsilon^{\mu\nu\rho\sigma}\; \text{Tr}\Bigl[\Big\{ \left(\partial_\mu \mathcal A_{\nu L}\right) \mathcal A_{\rho L}+\mathcal A_{\mu L}\left(\partial_\nu \mathcal A_{\rho L}\right)\Big\}\; U\mathcal A_{\sigma R}U^\dagger \nonumber \\ 
    &+\Big\{ \left(\partial_\mu \mathcal A_{\nu R}\right) \mathcal A_{\rho R}+\mathcal A_{\mu R}\left(\partial_\nu \mathcal A_{\rho R}\right)\Big\}\; U^\dagger \mathcal A_{\sigma L}U\Bigl]\;.
\end{align}
However, the introduction of background gauge fields is rather subtle since vector currents might not remain conserved because of mixed anomalies. Therefore, one needs to add new counterterms to maintain gauge invariance. Ref.~\cite{Harvey:2007ca} computed such terms for the first time and the relevant interactions look like:
\begin{align}
    S_c\left(A_{L,R},B_{L,R}\right) \supset \frac{N_C}{24\pi^2} \int d^4 x\; \epsilon^{\mu\nu\rho\sigma}\;\text{Tr}\Bigl[\Bigl\{\left(\partial_\mu A_{\nu L}\right)A_{\rho L}+A_{\mu L}\left(\partial_\nu A_{\rho L}\right)\Bigr\}B_{\sigma L}+\left(L\leftrightarrow R\right)\Bigr]\;.
\end{align}
The remaining part is straightforward, where we take into account the full action $S_{\text{WZW}}+S_c$ and use the following relations to obtain interactions between neutral mesons and SM gauge bosons:
\begin{align}
   & A_L = g_2 W^a \frac{\tau^a}{2} + g_1 W^0\; \text{diag}(1/6,1/6)\;, \; \; \; A_R = g_1 W^0\; \text{diag}(2/3,-1/3)\;, \nonumber \\
   & \hspace{3cm} B_V \equiv  \frac{2g_\omega}{3}\; \text{diag}(\omega,\omega)\;, \; \; \; B_A =0\;.
\end{align}
Summing over all three generations of quarks and leptons, we finally get 
\begin{equation}
    \mathcal L_{\text{WZW}} \supset \frac{N_C}{48\pi^2}\; g_2^2 g_\omega \tan\theta_W\; \epsilon^{\mu\nu\rho\sigma}\; F_{\mu\nu}\omega_\rho Z_\sigma\;+...
    \label{eq:WZW_Lag1}
\end{equation}
Obviously, we would generate a plethora of interactions involving charged and neutral background fields such as $\rho, a$, etc., and SM gauge fields. However, we consider the most dominant one as given in Eq.~(\ref{eq:WZW_Lag1}) for our analysis.

\bibliographystyle{JHEP}
\bibliography{bibliography}
\end{document}